\documentclass[11pt, a4paper]{article}
\usepackage{}

\usepackage{multirow}
\usepackage{indentfirst}
\usepackage{lineno}
\usepackage{graphicx}
\usepackage{ae}
\usepackage{amsmath}
\usepackage{amssymb}
\usepackage{latexsym}
\usepackage{url}
\usepackage{epsfig}
\usepackage{cite}
\usepackage{mathrsfs}
\usepackage{amsfonts}
\usepackage{amsthm}
\usepackage{float}
\usepackage{booktabs}
\usepackage[usenames,dvipsnames]{color}
\usepackage{subfigure}

\usepackage{color}

\long\def\delete#1{}

\newcommand{\be}{\begin{equation}}
\newcommand{\ee}{\end{equation}}
\newcommand{\ben}{\begin{equation*}}
\newcommand{\een}{\end{equation*}}
\newcommand{\bea}{\begin{eqnarray}}
\newcommand{\eea}{\end{eqnarray}}
\newcommand{\bean}{\begin{eqnarray*}}
\newcommand{\eean}{\end{eqnarray*}}

\newtheorem{thm}{Theorem}[section]

\newtheorem{defn}[thm]{Definition}

\numberwithin{equation}{section}

\title{A novel method based on node's correlation to evaluate the important nodes in complex networks \thanks{Supported by the National Natural Science Foundation of China (No.11361033) and the National Natural Science Foundation of China (No.11861045).}}

\author{Pengli Lu$^{1}$\thanks{Corresponding author. E-mail addresses: lupengli88@163.com (\textbf{P. Lu}), dongchen199508@163.com (\textbf{C. Dong}), gyh7001@163.com (\textbf{Y. Guo}).}, Chen Dong$^{1}$ \;and\; Yuhong Guo$^{2}$
\\
\footnotesize{1. School of Computer and Communication, Lanzhou University of Technology, Lanzhou, 730050, Gansu, China}\\
\footnotesize{2. School of Mathematics and Statistics, Hexi University, Zhangye, 734000, Gansu, China}}

\date{}

\begin{document}

\openup 0.5\jot
\maketitle

\noindent\textbf{Abstract: }Finding the important nodes in complex networks by topological structure is of great significance to network invulnerability. Several centrality measures have
been proposed recently to evaluate the performance of nodes based on their correlation, showing that the interaction between nodes has an influence on the
importance of nodes. In this paper, a novel method based on node's distribution and global influence in complex networks is proposed. Our main idea is that the
importance of nodes being linked not only to the relative position in the network but also to the correlations with each other. The nodes in the complex networks are classified according to the distance matrix, then the correlation coefficient between pairs of nodes is calculated. From the whole perspective in the network, the global similarity centrality ($GSC$) is proposed based on the relevance and shortest distance between any two nodes. The efficiency, accuracy and monotonicity of the proposed method are analyzed in two artificial datasets and eight real datasets of different sizes. Experimental results show that the performance of $GSC$ method outperforms those current state-of-the-art algorithms.


$\mathbf{0 Introduction}$

Complex system can be modeled or mapped as complex network structure consisting of nodes and edges, with every vertex represents an entity and edges denote the
relationships between pairs of entities. The identification of influential nodes has attracted many researchers in large and complex networks including social network, protein network, transportation network, information network and next generation network. If the influential nodes in a traffic network or protein network lose efficacy, the entire network may occur a catastrophic failure. In social network, information network and communication network, message can be spread easily and quickly throughout the network by influential nodes \cite{1,2}. The variety of users' needs leads to the discrepancy in information transmission efficiency, so it is impossible for all the information spread in time. The users on the corners always receive messages relatively late, which is meaningless to them \cite{3,4,5}.

In complex networks, finding the influential nodes which are willing to spread information is of great significance. News spreading starts from one or few users and the information diffuses to friends who are closely related to or interested in it, then these friends transmit the news to theirs friendship networks. A organization in social networks corresponds to a group of individuals with the same or similar backgrounds \cite{6,7,8}. Take the gymnasium as an example. Keep the store's management philosophy remains the same, the owners replace will only affect the employees of the gym, not the members. Therefore, these news will merely generate a great response among the employees rather than cause waves among the customers.

Influential users play an important role in the information spreading and ranking them according to theirs influence capability have received much attention in recent years. In order to find key nodes, researchers have proposed a number of centrality measures from different perspective. The most common ways are degree centrality which only considers the node's own topological structure \cite{9}, betweenness centrality and closeness centrality of the shortest distance between nodes \cite{10,11}, and k-core decomposition centrality about the relative position of nodes in the network \cite{12}. However, degree centrality lacks accuracy, betweenness centrality and closeness centrality are not applicable to large-scale networks, and k-core decomposition centrality tends to assign nodes with different spreading capability to the same k-shell index. Therefore, these existing methods have been proved not to meet the current needs \cite{13}. Local dimension centrality ($LD$) \cite{14} broken through the traditional global dimension thought pattern, which combined with the characteristics of the power low distribution of BA scale-free network and each node's attribute. The main idea behind the method is that the distribution concentration of remaining nodes is related to the position of the initial node. However, $LD$ centrality considers the node's influence range but neglects the correlation between pairs of nodes. Motivated by $LD$ centrality, we propose our method. The nodes in the network are classified by distance matrix and the pertinence between any two nodes is calculated by pearson correlation coefficient. From the global perspective of the network, the influence of the shortest distance and the correlation between any pairs of nodes on the importance of nodes are analyzed, and global similarity centrality ($GSC$) is proposed. In this paper, we apply the proposed method to different sizes networks and compare it with the state-of-the-art algorithms. Experiment results show that the proposed method has better performance in efficiency, accuracy and monotonicity than other popular measures.

The rest of the paper is organized as follows. Section 2 analyzes the existing methods of the node importance research. In Section 3, $GSC$ algorithm is
introduced. Experimental results and discussions are included in Section 4. Finally, conclusion of the paper is in Section 5.

\section{Related work}\label{Se2}
In this section, we will briefly introduce the current progress of identifying important nodes in complex networks. A series of classic centrality measures have
been proposed to evaluate the spreading capability of nodes. Degree centrality is a simple and straightforward way to measure the importance of nodes by counting the number of neighbors \cite{9}. However, there is a huge flaw in this measure. Simply figuring up the neighboring nodes' numbers but ignoring the importance of the vertices themselves may be result in nodes with smaller degree being more vital than larger ones. Otherwise, the relative position of nodes in complex networks is also a significant thought pattern. Compared with nodes in larger degree, smaller nodes are more likely to be in the key position of news spreading and play an important role in the whole network. This phenomena is also the starting point of betweenness centrality and closeness centrality \cite{10,11}. Based on the definition of h-index and the degree of each node's neighboring nodes, T. Zhou et al. proposed a more feasible evaluation measure of node importance than degree centrality \cite{15}. Considering the neighboring nodes performance can improve the accuracy of identifying important nodes, Q. Liu et al. proposed the local H-index centrality to promote the reliability of the measure \cite{16}. P.L. Lu et al. also proposed an extended H-index centrality based on local H-index centrality and clustering coefficient \cite{17}.

Besides, it is also an important topic to measure the importance of nodes by decomposing the network. Kitstak et al. proposed the $k$-core decomposition centrality ($KS$) to determine the importance of nodes based on their relative positions in the network \cite{12}. First, set the $KS$ value of all nodes in the network to 1, then find out all nodes with degree 1 in the network as well as remove these nodes and theirs edge relationships. Whereafter, recalculate the degree of nodes in the network, then delete the nodes with degree 1 and theirs edge relationships up to there are no nodes with degree 1 in the network. At this time, the $KS$ value of the remaining nodes in the network is set to 2, then the above operation is repeated through to there are no nodes with degree 2 in the network. So on until the network is completely decomposed or there are only isolated nodes. The larger the $KS$ value of a node, the closer it is to the center of the network. Considering the influence of neighboring nodes, J. Wang et al. proposed the neighborhood coreness centrality ($cn$), which reflected the relative distance between neighboring nodes and network center \cite{18}. In $k$-core decomposition centrality, the number of nodes deleted during each step also can reflect the performance of nodes. Mixed degree decomposition ($MDD$) considers the variation of network topology structure in each decomposing step \cite{19}. Qi et al. applied laplacian matrix and quasi-laplacian matrix to the study of node centrality in complex networks with the knowledge of graph theory, the importance of nodes was represented by calculating the change of spectral energy with nodes deletion, which greatly improved the practicability of the method \cite{20,21}.
For the first time, newton's classical mechanics theorem in physics is combined with the topological structure of complex networks to propose the newton gravity
centrality ($G$). The degree of nodes is corresponding to the mass of planets and the shortest distance between nodes is parallelism to the radius \cite{22}. Wang et al. proposed an improved newton gravity centrality ($IGC$), which replaced the degree of the node to the $k$-core \cite{23}. A. Namtirtha et al. further improved the newton gravity centrality and put forward a new idea, which combined the degree and core of nodes to evaluate node's importance \cite{24}. A. Dutta et al. analyzed the applicable network of degree centrality and $k$-core decomposition centrality, then combined these two measures and proposed a new method which can be applicable to different networks \cite{25}.

In addition to considering the spreading capability of one node, evaluating the importance of nodes from the network global perspective is also a widely used
measure. On the basis of kirchhoff polynomials, Z. Dai et al. proposed a spanning tree centrality method to determine important nodes and changed the evaluation
of node importance from simple networks to weighted networks \cite{26}. On this basis, a near-linear time algorithm based on kirchhoff index is proposed to
measure the edge centrality of weighted networks, which further broadens the application range of the algorithm \cite{27}. In combination with the basic
concept of fractal dimension in physics, Silva et al. proposed local dimension centrality to explore the nature of networks. Since each node in the network has
different sphere of influence, the local dimension will also change with the diverse of the central node, which has an impact on the feasibility of the method.
Therefore, Y. Deng et al. improved the local dimension centrality to make the method more practicable \cite{14}. Our method is proposed based on the shortest
distance and correlation between nodes to identify the importance of nodes more accurately.

\section{Proposed method}\label{Se3}
Distance matrix indicates the shortest distance between node pairs in the network, and it reflects the relative position of nodes. Core nodes are located at the center of the network, and the shortest path between many node pairs will go through these nodes, therefore the shortest distance between these nodes and other nodes is relatively small. Common nodes are located at the nooks of the network, while the surrounding nodes are dispersed, so the length of shorted paths are relatively large. Local dimension centrality ($LD$) combines the characteristics of distance matrix with the power law distribution, which matches the importance of nodes with the scale of locality of each node. The lower $LD$ means the higher importance. In other words, the distance between the node and the core of the network also affects the importance of the node, and nodes in the dense location are often more important than nodes in the sparse location. However, the local dimension centrality only considers the distribution of nodes and does not take the properties of vertices as the evaluation criterion. Therefore, an accurate algorithm considering node's property is certainly needed.

Let $G=(V,E)$ be an unweighted network with vertex set $V(G)=\{ 1, 2, 3, .., N \}$ and edge set $E(G)$. We define the weighted matrix $W(G)$ of size $N \times N$ as follows:
    \begin{equation}\label{adjmatix}
    \begin{aligned}
    \begin{split}
    W(G)=
    \begin{cases}
    0,&\mbox{if i=j}\\
    1,&\mbox{if i and j are adjacent}\\
    \infty,&\mbox{if i and j are not adjacent}\\
    \end{cases}
    \end{split}
    \end{aligned}
    \end{equation}

The distance between two nodes $i,j \in V(G)$, denoted by $d_{i,j}$, is the length of the shortest path from node $i$ to $j$. The distance matrix of $G$,
denoted by $D(G)$, is a $N \times N$ matrix with the $(i,j)-$th entry being $d_{i,j}$, defined as follows:
    \begin{equation}\label{adjmatix}
    \begin{aligned}
    \begin{split}
    D(G)=
        \begin{bmatrix}
        d_{1,1} & d_{1,2} & ... & d_{1,N}\\
        d_{2,1} & d_{2,2} & ... & d_{2,N}\\
        ... & ... & ... & ...\\
        d_{N,1} & d_{N,2} & ... & d_{N,N}\\
        \end{bmatrix}
    \end{split}
    \end{aligned}
    \end{equation}

Distance matrix of the network can be obtained by calculating the two-node shortest distance from $W(G)$ by Floyd-Warshall algorithm.

The maximus distance from node $i$ to other nodes, which represents the surrounding size of node $i$, is denoted as:
    \begin{equation}\label{adjmatix}
    \begin{aligned}
    \begin{split}
    D_{i}=max(d_{i,j}),~j \in V
    \end{split}
    \end{aligned}
    \end{equation}
and the diameter $D$ of the network is:
    \begin{equation}\label{adjmatix}
    \begin{aligned}
    \begin{split}
    D=max(D_{i})
    \end{split}
    \end{aligned}
    \end{equation}

After calculating the relative distance between nodes, node distribution vector and distance vector are defined based on the location of each node.
\begin{defn}
{
(Node Distribution Vector and Distance Vector)
\em
The node distribution vector $NDV_{i}$ and distance vector $DV_{i}$ for node $i$ are defined as follows, where $|V_{i}^{k}|$ represents the number of nodes in the network whose shortest distance from node $i$ is $k$.
    \begin{equation}\label{adjmatix}
    \begin{aligned}
    \begin{split}
    NDV_{i}=( |V_{i}^{1}|, |V_{i}^{2}|, |V_{i}^{3}|, ..., |V_{i}^{D}| )
    \end{split}
    \end{aligned}
    \end{equation}

    \begin{equation}\label{adjmatix}
    \begin{aligned}
    \begin{split}
    DV_{i}=( |V_{i}^{1}|, 2|V_{i}^{2}|, 3|V_{i}^{3}|, ..., D|V_{i}^{D}| )
    \end{split}
    \end{aligned}
    \end{equation}
}
\end{defn}

    \begin{figure}
        \centering
        \includegraphics[width=8cm]{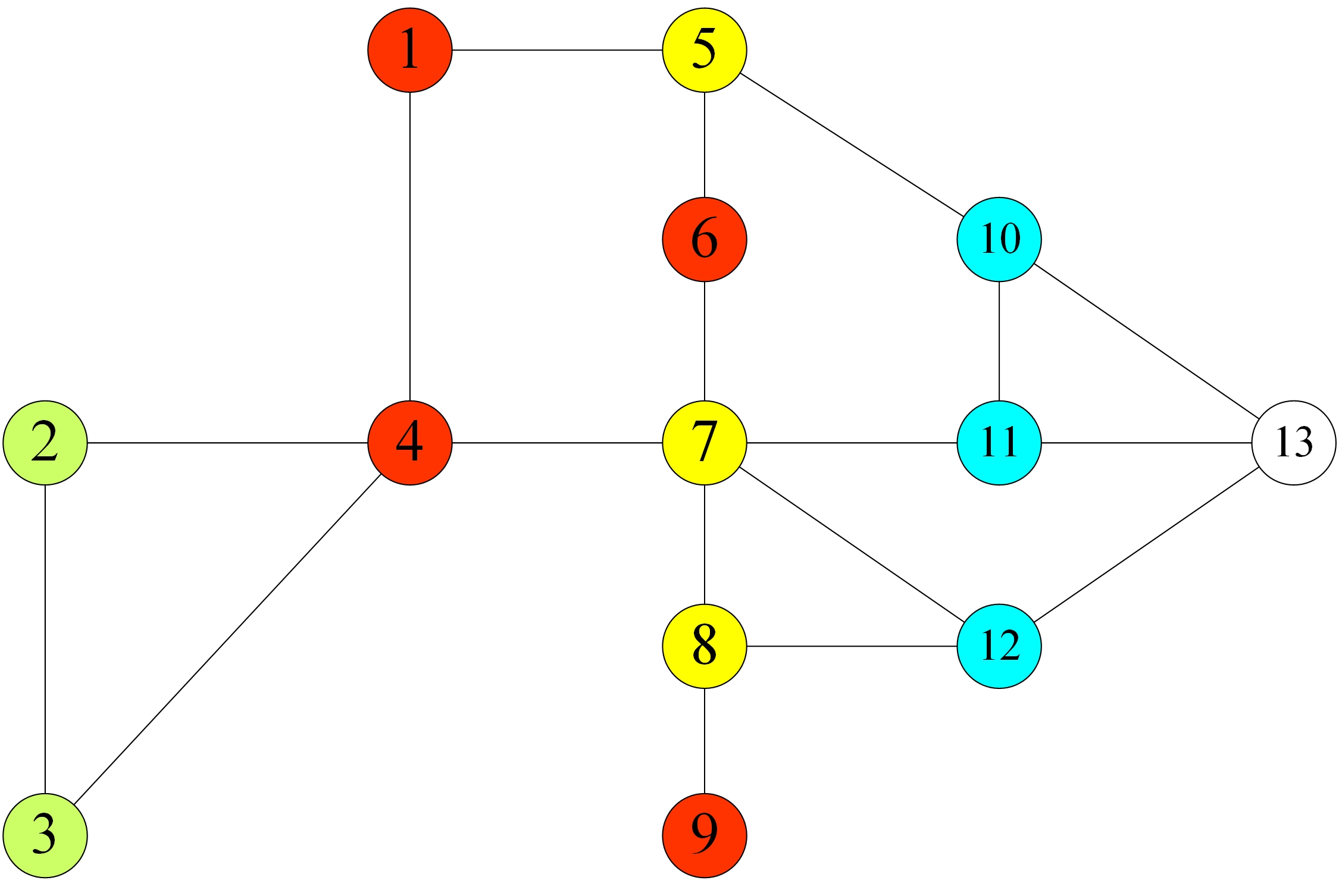}\\
        \caption{A simple graph. (Take node 13 as the initial node, the nodes are divided into four parts by the distance from node 13 and each shown in different colors)}\label{1}
    \end{figure}

As shown in Fig. 1, we take node 13 as the initial node and divide the other nodes in the network into four levels. The distance between nodes 10, 11, 12 and 13 are 1, the distance between nodes 5, 7, 8 and 13 are 2, the distance between nodes 1, 4, 6, 9 and 13 are 3, the distance between nodes 2, 3 and 13 are 4, and $D_{13}$ is 4. We can represent the distribution vector of node 13 as $NDV_{13}=(3, 3, 4, 2)$. Node 8 has divided nodes into 3 levels, hence $NDV_{8}=(3, 4, 5, 0)$. Node 7 has divided nodes into 2 levels and $NDV_{7}=(5, 7, 0, 0)$. The relative position of nodes in the network has an obvious impact on the distribution vector. The closer to the center, the denser the distribution vector of node is. While the nodes on the corners are relatively dispersed. The distance vector for nodes 7, 8, 13 are: $DV_{7}= (5, 14, 0, 0)$, $DV_{8}= (3, 8, 15, 0)$, $DV_{13}= (3, 6, 12, 8)$, respectively.

After obtaining the distribution vector and distance vector of nodes, pearson correlation coefficient is used to measure the commonality between two vectors, which are defined as $P_{i,j}$ and $D_{i,j}$, respectively. $P_{i,j}$ describes the similarity between nodes in node distribution and is calculated using the traditional pearson correlation coefficient formula, while $D_{i,j}$ improves pearson correlation coefficient according to the distance distribution of nodes, which calculates the correlation between the distance distribution and the average shortest distance of nodes. By counting the number of nodes on each distance and calculating the difference between the distance of two nodes and the average shortest distance, the similarity of topological structure between pairs of nodes is reflected and the relative position of nodes in the network can be expressed. The specific formulas are as follows:
    \begin{equation}\label{adjmatix}
    \begin{aligned}
    \begin{split}
    P_{i,j}=\frac{\displaystyle \sum_{k=1}^ {D}( NDV_{i}^{k}- \overline{NDV_{i}} ) \cdot ( NDV_{j}^{k}- \overline{NDV_{j}} )}
    {\sqrt{\displaystyle \sum_{k=1}^ {D} ( NDV_{i}^{k}- \overline{NDV_{i}} )^{2}} \cdot \sqrt{\displaystyle \sum_{k=1}^ {D} ( NDV_{j}^{k}- \overline{NDV_{j}} )^{2}}}
    \end{split}
    \end{aligned}
    \end{equation}

    \begin{scriptsize}
    \begin{equation}\label{adjmatix}
    \begin{aligned}
    \begin{split}
    D_{i,j}=\frac{\displaystyle \sum_{k=1}^ {D} \left[ NDV_{i}^{k} \times \left ( \frac{DV_{i}^{k}}{NDV_{i}^{k}}- (\overline{DV_{i}} \times \frac{D}{N-1}) \right) \right] \cdot \left [ NDV_{j}^{k} \times \left ( \frac{DV_{j}^{k}}{NDV_{j}^{k}}- (\overline{DV_{j}} \times \frac{D}{N-1}) \right) \right] }
    {\sqrt{\displaystyle \sum_{k=1}^ {D} \left[ NDV_{i}^{k} \times \left ( \frac{DV_{i}^{k}}{NDV_{i}^{k}}- (\overline{DV_{i}} \times \frac{D}{N-1}) \right) \right ] ^{2}} \cdot \sqrt{\displaystyle \sum_{k=1}^ {D} \left [ NDV_{j}^{k} \times \left ( \frac{DV_{j}^{k}}{NDV_{j}^{k}}- (\overline{DV_{j}} \times \frac{D}{N-1}) \right) \right ] ^{2}}}
    \end{split}
    \end{aligned}
    \end{equation}
    \end{scriptsize}
where $NDV_{i}^{k}$ denotes the $k-$th element value of vector $NDV_{i}$, and $\overline{NDV_{i}}$ is the mean value of the vector $NDV_{i}$, $DV_{i}^{k}$ and $\overline{DV_{i}}$ are represented as the element value and mean value of vector $DV_{i}$. The results of Eq.(3.7) and Eq.(3.8) are between $-1$ and $+1$, greater than 0 means positive correlation between two vectors, less than 0 means negative correlation between two vectors, and equal to 0 means there is no correlation. Take Fig. 1 as an example, the correlation coefficient between node 7 and node 8 are\\
$P_{7,8}=\frac{\displaystyle \sum_{k=1}^ {D}( NDV_{7}^{k}- \overline{NDV_{7}} ) \cdot ( NDV_{8}^{k}- \overline{NDV_{8}} )}
{\sqrt{\displaystyle \sum_{k=1}^ {D} ( NDV_{7}^{k}- \overline{NDV_{7}} )^{2}} \cdot \sqrt{\displaystyle \sum_{k=1}^ {D} ( NDV_{8}^{k}- \overline{NDV_{8}} )^{2}}} \cong 0.3035$ and\\
\begin{footnotesize}
$D_{7,8}=\frac{\displaystyle \sum_{k=1}^ {D} \left[ NDV_{7}^{k} \times \left ( \frac{DV_{7}^{k}}{NDV_{7}^{k}}- (\overline{DV_{7}} \times \frac{D}{N-1}) \right) \right] \cdot \left [ NDV_{8}^{k} \times \left ( \frac{DV_{8}^{k}}{NDV_{8}^{k}}- (\overline{DV_{8}} \times \frac{D}{N-1}) \right) \right] }
    {\sqrt{\displaystyle \sum_{k=1}^ {D} \left[ NDV_{7}^{k} \times \left ( \frac{DV_{7}^{k}}{NDV_{7}^{k}}- (\overline{DV_{7}} \times \frac{D}{N-1}) \right) \right ] ^{2}} \cdot \sqrt{\displaystyle \sum_{k=1}^ {D} \left [ NDV_{8}^{k} \times \left ( \frac{DV_{8}^{k}}{NDV_{8}^{k}}- (\overline{DV_{8}} \times \frac{D}{N-1}) \right) \right ] ^{2}}} \cong 0.7069$,\\
\end{footnotesize}
the correlation coefficient between node 7 and node 13 are \\
$P_{7,13}=\frac{\displaystyle \sum_{k=1}^ {D}( NDV_{7}^{k}- \overline{NDV_{7}} ) \cdot ( NDV_{13}^{k}- \overline{NDV_{13}} )}
{\sqrt{\displaystyle \sum_{k=1}^ {D} ( NDV_{7}^{k}- \overline{NDV_{7}} )^{2}} \cdot \sqrt{\displaystyle \sum_{k=1}^ {D} ( NDV_{13}^{k}- \overline{NDV_{13}} )^{2}}}=0$ and\\
\begin{footnotesize}
$D_{7,13}=\frac{\displaystyle \sum_{k=1}^ {D} \left[ NDV_{7}^{k} \times \left ( \frac{DV_{7}^{k}}{NDV_{7}^{k}}- (\overline{DV_{7}} \times \frac{D}{N-1}) \right) \right] \cdot \left [ NDV_{13}^{k} \times \left ( \frac{DV_{13}^{k}}{NDV_{8}^{k}}- (\overline{DV_{13}} \times \frac{D}{N-1}) \right) \right] }
    {\sqrt{\displaystyle \sum_{k=1}^ {D} \left[ NDV_{7}^{k} \times \left ( \frac{DV_{7}^{k}}{NDV_{7}^{k}}- (\overline{DV_{7}} \times \frac{D}{N-1}) \right) \right ] ^{2}} \cdot \sqrt{\displaystyle \sum_{k=1}^ {D} \left [ NDV_{13}^{k} \times \left ( \frac{DV_{13}^{k}}{NDV_{13}^{k}}- (\overline{DV_{13}} \times \frac{D}{N-1}) \right) \right ] ^{2}}} \cong 0.6789$,\\
\end{footnotesize}
thus, node 8 plays more active role in news spreading of node 7 in the network than node 13.

\begin{defn}
{
(Global Similarity Centrality)
\em
The global similarity centrality consists of two parts, and it is defined as follows:
    \begin{equation}\label{adjmatix}
    \begin{aligned}
    \begin{split}
    NC_{i,j}=
    \left\{
    \begin{aligned}
    \frac{1-P_{i,j}}{d_{i,j}}+(1+\frac{D_{i,j}}{d_{i,j}}), P_{i,j}>0\\
    \frac{1+P_{i,j}}{d_{i,j}}+(1+\frac{D_{i,j}}{d_{i,j}}), P_{i,j}<0\\
    1+\frac{D_{i,j}}{d_{i,j}} ~~~~~~~~~~~~~~~~,P_{i,j}=0\\
    \end{aligned}
    \right.
    \end{split}
    \end{aligned}
    \end{equation}

    \begin{equation}\label{adjmatix}
    \begin{aligned}
    \begin{split}
    GSC_{i}=\displaystyle \sum_{v_{j} \in V} NC_{i,j}
    \end{split}
    \end{aligned}
    \end{equation}
}
\end{defn}

\begin{table}
\begin{center}
\begin{tabular}{l}
 \toprule
 Algorithm: Ranking nodes on the basis of cumulative centrality\\
 \midrule
 $\mathbf{01}$~~~~~~~~~~~~~~~~$\mathbf{Input}$:~~~~$G=(V,E)$\\
 $\mathbf{02}$~~~~~~~~~~~~~~~~$\mathbf{Output}$:~~A ranking list of nodes' importance\\
 $\mathbf{03}$~~~~~~~~~~~~~~~~$\mathbf{Begin~~Algorithm}$\\
 $\mathbf{04}$~~~~~~~~~~~~~~~~~~~~~~Floyd-Warshall algorithm is used to calculate the shortest\\
              ~~~~~~~~~~~~~~~~~~~~~~~~distance between nodes and the diameter of the graph $G$\\
 $\mathbf{05}$~~~~~~~~~~~~~~~~~~~~~~$\mathbf{For}$~~i=1~~$\mathbf{to}$~~$|V|$\\
 $\mathbf{06}$~~~~~~~~~~~~~~~~~~~~~~~~~~~~Calculate $NDV_{i}$ and $DV_{i}$ using $Eq.(3.5)$ and $Eq.(3.6)$\\
 $\mathbf{07}$~~~~~~~~~~~~~~~~~~~~~~$\mathbf{End}$~~$\mathbf{for}$\\
 $\mathbf{08}$~~~~~~~~~~~~~~~~~~~~~~$\mathbf{For}$~~i=1~~$\mathbf{to}$~~$|V|$\\
 $\mathbf{09}$~~~~~~~~~~~~~~~~~~~~~~~~~~~~Set $GSC_{i}$=0\\
 $\mathbf{10}$~~~~~~~~~~~~~~~~~~~~~~~~~~~~$\mathbf{For}$~~j=1~~$\mathbf{to}$~~$|V|$\\
 $\mathbf{11}$~~~~~~~~~~~~~~~~~~~~~~~~~~~~~~~~~~Calculate $P_{i,j}$ and $D_{i,j}$ using $Eq.(3.7)$ and $Eq.(3.8)$\\
 $\mathbf{12}$~~~~~~~~~~~~~~~~~~~~~~~~~~~~~~~~~~According to the value of $P_{i,j}$ and $D_{i,j}$ to use $Eq.(3.9)$\\
              ~~~~~~~~~~~~~~~~~~~~~~~~~~~~~~~~~~calculate $NC_{i,j}$ \\
 $\mathbf{13}$~~~~~~~~~~~~~~~~~~~~~~~~~~~~~~~~~~$GSC_{i}=GSC_{i}+NC_{i,j}$\\
 $\mathbf{14}$~~~~~~~~~~~~~~~~~~~~~~~~~~~~$\mathbf{End}$~~$\mathbf{for}$\\
 $\mathbf{15}$~~~~~~~~~~~~~~~~~~~~~~$\mathbf{End}$~~$\mathbf{for}$\\
 $\mathbf{16}$~~~~~~~~~~~~~~~~~~~~~~Sort the nodes in descending order based on $GSC$ values to\\
             ~~~~~~~~~~~~~~~~~~~~~~~~obtain the ranking list\\
 $\mathbf{17}$~~~~~~~~~~~~~~~~$\mathbf{End~~Algorithm}$\\
 \bottomrule
\end{tabular}
\end{center}
\end{table}

The formula in this section consists of two parts: the distance clustering coefficient of nodes, and the global correlation of node $i$. When the distance
coefficient is minus, node $j$ has a negative effect on the spreading ability of node $i$, which affects the propogating of node $i$ in the network. Therefore,
$1+P_{i,j}$ is used to accurately calculate the clustering coefficient between node $i$ and node $j$. At the same time, considering the different influence
capability between nodes, coefficient $d_{i,j}$ and $D_{i,j}$ also have positive effect on the whole algorithm, and $1+\frac{D_{i,j}}{d_{i,j}}$ is used to control the influence of distance between nodes on the proposed method.

Algorithm provides an idea of the proposed method which contains specific calculation details of each step. Floyd-Warshall algorithm is used in line 4 to calculate the distance matrix and the diameter of graph $G$, lines 5-7 use $Eq.(3.5)$ and $Eq.(3.6)$ to calculate the node distribution vector for each node, the correlation coefficient of node $i$ and the other nodes in the network in lines 8-15 through formula $Eq.(3.7)$ and $Eq.(3.8)$ to calculate, then compute node's $GSC$. Finally, the nodes are sorted by the value of the $GSC$. The time complexity of the Floyd-Warshall algorithm is $O(|V|^{3})$, the rest of the proposed measure is $O(|V|+|E|)$.

When evaluating the importance of nodes, the proposed method first defines the distribution vector and distance vector of nodes according to the structure of the network, then calculates the similarity degree between pairs of nodes with pearson correlation coefficient and the importance of nodes are based on the node's correlation. Compared with the existing global clustering coefficient algorithm, the proposed algorithm has made the improvement on the clustering method. We consider the network structure, and also based on the similarity degree between nodes, re-divided the nodes from the perspective of propagation. The measure can determine the node's spreading capability more accurately, which make up the shortcomings of the global clustering coefficient algorithm for only considering the single parameter.

\section{Experimental results and discussions}

In this section, to evaluate the proposed method, we compare it with a series of currently popular algorithms, including: K-Shell decomposition centrality ($KS$) \cite{12}, neighborhood coreness centrality ($cn$) \cite{20}, H-index centrality ($H$) \cite{15}, Local H-index centrality ($LH$) \cite{16}, Newton's gravity centrality ($G$) \cite{24}, Improved Newton's gravity centrality ($IGC$) \cite{25}, K-shell hybrid method ($Ksh$) \cite{26}, Weighted k-shell degree neighborhood centrality ($Ksd$) \cite{27}, Betweenness centrality ($BC$) \cite{10}, Closeness centrality ($CC$) \cite{11}, Eigenvector centrality ($EC$) \cite{48} and Pagerank centrality ($PA$) \cite{49}. Then, these methods are used in eight real-world datasets and two artificial datasets. The networks used in this paper are all undirected networks, and the algorithms are not experimented in directed networks. Real-world datasets including network of mutual relations between club employees and customers (Karate) \cite{28},  Lusseau's Bottlenose Dolphins social network (Dolphins) \cite{29}, the network of selling political books about the presidential election in Amazon during 2004 (Polbooks) \cite{30}, the schedule network of major league soccer clubs (Football) \cite{31}, a network of collaborative relationships among jazz musicians (Jazz) \cite{32}, American airlines flight route network (USAir) \cite{33}, Rovira Virgili university E-mail message network between teachers and students (Email) \cite{34}, a network of interrelationships between proteins (Yeast) \cite{35}. In artificial network datasets, including Small-World network (WS) \cite{36} and Lancichinetti-Fortunato-Radicchi network (LFR-2000) \cite{37}, both sets of these datasets are generated by software Gephi. The specific parameters of the datasets are shown in Table 1.

\begin{table}[H]
Table 1: specific parameters of the datasets.

\begin{tabular}{cccccccc}
 \toprule
 Network & |N| & |E| & Average number & Maximum degree &  $\beta_{th}$  &  $\beta$  &  Assortativity\\
 \midrule
 Karate   & 34  & 78   & 4.588  & 17    & 0.129  & 0.13 & -0.4756\\
 Dolphins & 62  & 159  & 5.129  & 12    & 0.147  & 0.15 & -0.0436\\
 Polbooks & 105 & 441  & 8.400  & 25    & 0.0838 & 0.09 & -0.1279\\
 Football & 115 & 613  & 10.661 & 12    & 0.0932 & 0.10 &  0.1624\\
 Jazz     & 198 & 2742 & 27.967 & 100   & 0.026  & 0.03 &  0.0202\\
 USair    & 332 & 2126 & 12.81  & 139   & 0.0225 & 0.03 & -0.2079\\
 Email    & 1133 & 5451 & 9.622 & 71    & 0.0535 & 0.06 &  0.0782\\
 WS       & 2000 & 6012 & 6.021 & 11    & 0.1559 & 0.16 & -0.0563\\
 LFR-2000 & 2000 & 4997 & 9.988 & 39    & 0.0477 & 0.05 & -0.0032\\
 Yeast    & 2361 & 7181 & 6.083 & 65    & 0.0600 & 0.07 & -0.0489\\
 \bottomrule
\end{tabular}
\end{table}

\subsection{Discrimination capability}

In this experiment, we will study the discriminating ability of ranking lists generated by involved measures from the aspects of monotonicity and
resolution \cite{38,39}. In order to better evaluate the performance of nodes and calculate the capability of different measures to distinguish the importance of nodes, researchers applied monotonicity to assess the ability of different measures about distributing the spreading efficiency of nodes in social networks.
The formula for monotonicity is as follows:
    \begin{equation}\label{adjmatix}
    \begin{aligned}
    \begin{split}
    M(A)=\left (1-\frac{\sum_{a \in A} |X|_{a} \times (|X|_{a}-1)} {|X| \times (|X|-1)} \right)^{2}
    \end{split}
    \end{aligned}
    \end{equation}
where $A$ is the ranking list of one measure, $|X|$ is the total nodes number of $A$, $|X|_{a}$ is the number of nodes in level a. The range of monotonicity is
[0,1]. The better the measure's discrimination ability, the bigger the value of monotonicity is. Experimental results are shown in Table 2. Involved methods are
applied to different networks for comparison, the results show that the measure which considers the performance of neighboring nodes ($cn$, $LH$) can better reveal the discrimination ability of nodes than only a single node ($KS$, $H$), and the proposed method $GSC$ indicates the best performance while the existing algorithms $Ksd$, $BC$ and $EC$ also perform well.

In order to further compare the ability of different methods to distinguish node importance, the second part of the experiment uses the cumulative distribution function ($CDF$) curve to represent the resolution of these methods. $A$ represents the ranking list generated by one measure, while the $CDF$ of $A$ represents the probability that the element in $A$ is less than or equal to a given value. In other words, the slower the curve rises, the higher the resolution of the method, and the better it is to distinguish the importance of nodes. Fig. 2 compares the $CDF$ curves of the ranking list generated by different algorithms including $GSC$. Experimental results show that the proposed method has best performance in distinguishing node importance.

\begin{table}[H]
Table 2: The $M$ value of ranking list generated by different measures in different networks.

\tiny
\begin{tabular}{cccccccccccccc}
 \toprule
 Network & M(KS) & M(cn) & M(H) & M(LH) & M(G) & M(IGC) & M(Ksh) & M(Ksd) & M(BC) & M(CC) & M(EC) & M(PA) & M(GSC)\\
 \midrule
 Karate   & 0.4958 & 0.8526 & 0.5766 & 0.8925 & 0.9334 & 0.9577  & 0.9334 & 0.9542 & 0.7754 & 0.8993 & $\mathbf{0.9612}$ & 0.9542 & 0.9542\\
 Dolphins & 0.3769 & 0.9284 & 0.6841 & 0.9592 & 0.9916 & 0.9947  & 0.9937 & $\mathbf{0.9979}$ & 0.9623 & 0.9737 & $\mathbf{0.9979}$ & 0.9905 & $\mathbf{0.9979}$\\
 Polbooks & 0.4949 & 0.9641 & 0.7067 & 0.9821 & 0.9982 & 0.9993  & 0.9993 & $\mathbf{0.9999}$ & 0.9974 & 0.9847 & $\mathbf{0.9999}$ & 0.9903 & $\mathbf{0.9999}$\\
 Football & 0.0003 & 0.4218 & 0.2349 & 0.9190 & 0.8626 & 0.9903  & 0.8626 & 0.9994 & $\mathbf{0.9999}$ & 0.9488 & $\mathbf{0.9999}$ & 0.9903 & $\mathbf{0.9999}$\\
 Jazz     & 0.7944 & 0.9982 & 0.9383 & 0.9982 & 0.9995 & 0.9995  & $\mathbf{0.9996}$ & 0.9995 & 0.9885 & 0.9878 & 0.9659 & 0.9993 & $\mathbf{0.9996}$\\
 USair    & 0.8114 & 0.9628 & 0.8335 & 0.9856 & 0.9942 & 0.9949  & 0.9943 & $\mathbf{0.9951}$ & 0.6970 & 0.9892 & 0.9943 & 0.9943 & $\mathbf{0.9951}$\\
 Email    & 0.8089 & 0.9839 & 0.8584 & 0.9899 & 0.9996 & 0.9998  & $\mathbf{0.9999}$ & $\mathbf{0.9999}$ & 0.9400 & 0.9988 & 0.8875 & 0.9988 & $\mathbf{0.9999}$\\
 WS       & 0.0002 & 0.6085 & 0.2904 & 0.9155 & 0.9757 & 0.9982  & 0.9799 & 0.9998 & $\mathbf{0.9999}$ & 0.9987 & $\mathbf{0.9999}$ & 0.9954 & $\mathbf{0.9999}$\\
 LFR-2000 & 0.0385 & 0.9789 & 0.7184 & 0.9927 & 0.9997 & 0.9998  & 0.9998 & $\mathbf{0.9999}$ & $\mathbf{0.9999}$ & 0.9951 & 0.5618 & 0.7242 & $\mathbf{0.9999}$\\
 Yeast    & 0.6643 & 0.9458 & 0.6873 & 0.9686 & 0.9959 & 0.9964  & 0.9963 & 0.9964 & 0.7012 & 0.9964 & 0.7210 & 0.9916 &$\mathbf{0.9965}$\\
 \bottomrule
\end{tabular}
\end{table}

    \begin{figure}
        \centering
        \includegraphics[width=18cm]{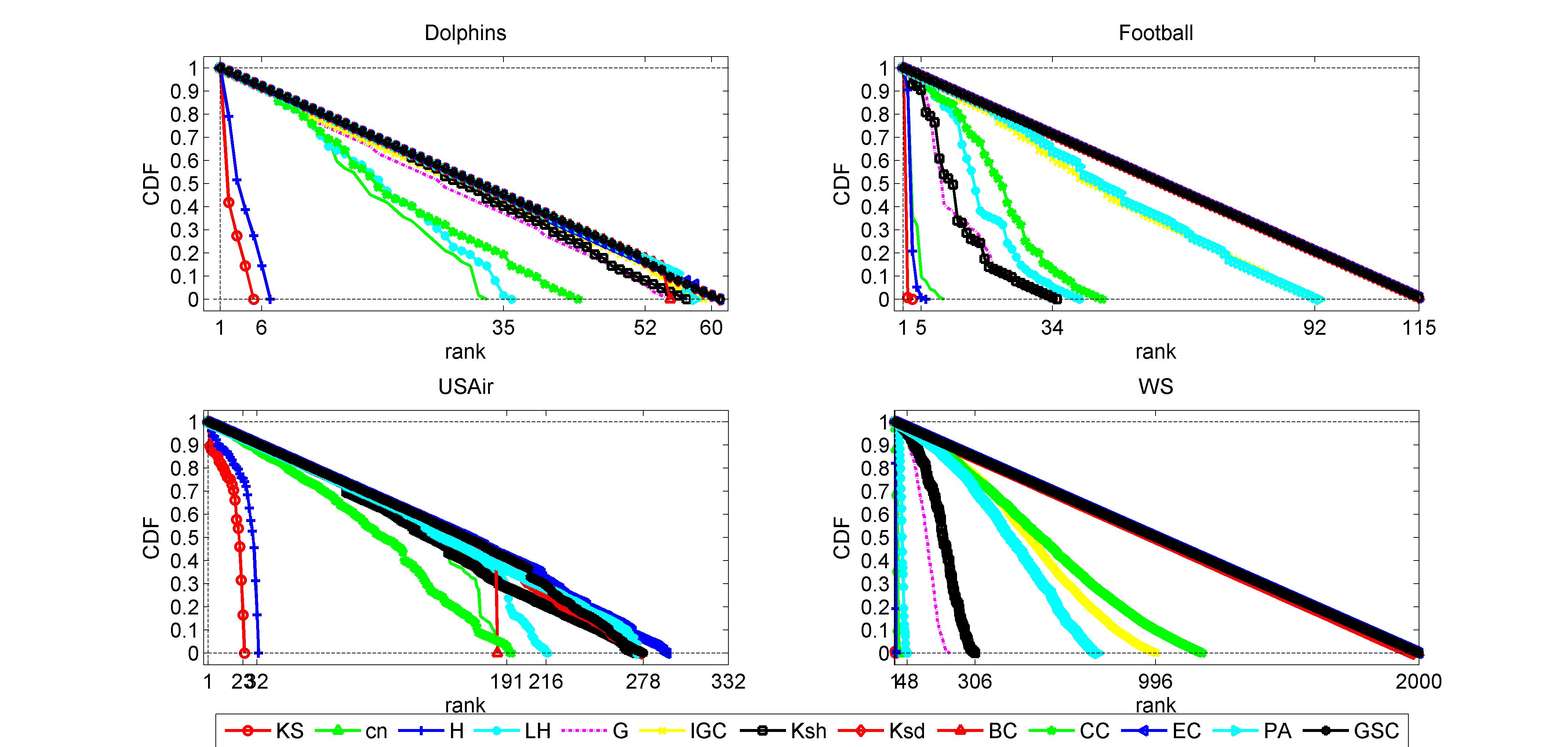}\\
        \caption{The $CDF$ curve of all measures on Dolphins, Football, USAir, WS networks.}\label{1}
    \end{figure}

\subsection{Accuracy of spreading model}
In this experiment, we will compare the accuracy between the ranking lists obtained by different measures and the real spreading capability of nodes. In order to acquire the performance of nodes, we simulated the spreading process of nodes in the traditional epidemic spreading model, then calculated the correlation between the results and ranking lists obtained by different algorithms. Susceptible-Infected-Recovered ($SIR$) model has become the most popular epidemic spreading model because of its simple principle and wide range of applications, it also has been applied to different articles \cite{40,41,42,43}.

In standard $SIR$ model, every node has only three different states: susceptibility ($S$), infection ($I$), and recovery ($R$). In order to obtain the spreading capability of each node, we only set one node to the infected state at the beginning of the experiment, while all the remaining nodes are set to the susceptible state. In each time period, the infected nodes will spread to all the susceptible nodes which connected to them with probability $\alpha$, and these nodes will also recover with probability $\beta$ after being infected. After the experiment, the number of nodes in the recovery state is defined as the real spreading capability of the nodes. The above experiment is repeated for 1000 times, so that all nodes of the network can obtain the spreading capacity range of nodes and take the average value as the final result. The threshold of the network is defined as $\beta_{th}=\frac{<d>}{<d>^{2}}$, where $<d>$ is the average degree of the node, and $<d>^{2}$ is the average degree of the second-order neighbors of the node. The threshold $\beta_{th}$ and the corresponding $\beta$ are shown in Table 1.

After obtaining the spreading capability of nodes, we use Kendall correlation coefficient to evaluate the relativity between the ranking lists obtained by
different algorithms and the real spreading capability \cite{44,45,46}. Let $X$ and $Y$ be two sets of the ranking sequences, and $(x_{1}, y_{1}), (x_{2}, y_{2}),$ $..., (x_{n}, y_{n})$ be a set of the ranking pairs. Two data pairs $(x_{i}, y_{i})$ and $(x_{j}, y_{j})$ are considered to be concordant under the condition that if ($x_{i} > x_{j}$ and $y_{i} > y_{j}$) or ($x_{i} < x_{j}$ and $y_{i} < y_{j}$), and discordant under the condition that if ($x_{i} > x_{j}$ and $y_{i} < y_{j})$ or ($x_{i} < x_{j}$ and $y_{i} > y_{j}$). The Kendall correlation coefficient is defined as follows
    \begin{equation}\label{adjmatix}
    \begin{aligned}
    \begin{split}
    \tau = \frac{2(R_{a}-R_{b})}{R(R-1)}
    \end{split}
    \end{aligned}
    \end{equation}
where $R_{a}$, $R_{b}$ are the numbers of concordant and discordant pairs, $n$ is the number of all pairs.

Table 3 shows the correlation at a certain point between the node's real spreading capability and ranking lists generated by involved algorithms. It is obvious that the proposed measure $GSC$ has the best performance in 9 of the 10 experimental datasets, while the $cn$ algorithm has the best performance in Football network, and $KS$, $BC$ and $PA$ show the worst effect in all networks due to the limitations of the algorithm. These results reflect the superiority of the proposed method over the other state-of-the-art algorithms.

\begin{table}[H]
Table 3: The kendall $\tau$ value of each method in 10 networks with a given $\beta$ value.

\scriptsize
 \begin{tabular}{cccccccccccccc}
 \toprule
 Network & KS & cn & H & LH & G & IGC & Ksh & Ksd & BC & CC & EC & PA & GSC\\
 \midrule
 Karate   & 0.5799 & 0.6789 & 0.6219 & 0.7079 & 0.7580 & 0.7838 & 0.7472 & 0.7972 & 0.5433 & 0.6626 & 0.8245 & 0.3535 & $\mathbf{0.8332}$\\
 Dolphins & 0.7363 & 0.8275 & 0.8420 & 0.8678 & 0.7499 & 0.8091 & 0.5810 & 0.7984 & 0.5900 & 0.6175 & 0.6132 & 0.5948 & $\mathbf{0.9006}$\\
 Polbooks & 0.7196 & 0.8143 & 0.7946 & 0.8507 & 0.7505 & 0.7713 & 0.6196 & 0.7628 & 0.3646 & 0.3715 & 0.5818 & 0.4516 & $\mathbf{0.8693}$\\
 Football & 0.1320 & 0.4931 & 0.3897 & $\mathbf{0.4453}$ & 0.4127 & 0.3945 & 0.3220 & 0.3997 & 0.1246 & 0.1522 & 0.3475 & 0.3079 & 0.4235\\
 Jazz     & 0.7690 & 0.8765 & 0.8615 & 0.8885 & 0.8001 & 0.8102 & 0.7228 & 0.8344 & 0.4912 & 0.7219 & 0.8458 & 0.5949 & $\mathbf{0.8909}$\\
 USair    & 0.7550 & 0.8462 & 0.7580 & 0.8478 & 0.7532 & 0.7782 & 0.4633 & 0.8232 & 0.5590 & 0.7805 & 0.8361 & 0.3710 & $\mathbf{0.8851}$\\
 Email    & 0.8218 & 0.8631 & 0.8401 & 0.8840 & 0.8359 & 0.8533 & 0.6854 & 0.8161 & 0.8210 & 0.8190 & 0.8517 & 0.5747 & $\mathbf{0.8872}$\\
 WS       & 0.1239 & 0.6701 & 0.5227 & 0.6515 & 0.6255 & 0.6384 & 0.4932 & 0.6373 & 0.6052 & 0.5872 & 0.6235 & 0.4657 & $\mathbf{0.7140}$\\
 LFR-2000 & 0.4049 & 0.7004 & 0.6795 & 0.7065 & 0.6614 & 0.6571 & 0.5360 & 0.6811 & 0.6843 & 0.7033 & 0.7157 & 0.6278 & $\mathbf{0.7090}$\\
 Yeast    & 0.7553 & 0.8231 & 0.7604 & 0.8492 & 0.7983 & 0.8108 & 0.5835 & 0.7703 & 0.6301 & 0.5653 & 0.7270 & 0.3046 & $\mathbf{0.8686}$\\
 \bottomrule
\end{tabular}
\end{table}

Otherwise, we research the accuracy of the algorithm in the $SIR$ model under different infection rates. Taking four networks of different sizes as an example, Fig. 3 expresses the correlation curve between the ranking lists and real spreading ability of nodes. In the experiment networks, with the increasing of $\beta$, the proposed method is more accurate than other methods. Especially near the threshold $\beta_{th}$, the accuracy reaches the peak. The performance of the existing algorithms is equal to $GSC$ measure in the comparison of discriminating ability, while these measures are far less than $GSC$ in accuracy.

    \begin{figure}
        \centering
        \includegraphics[width=18cm]{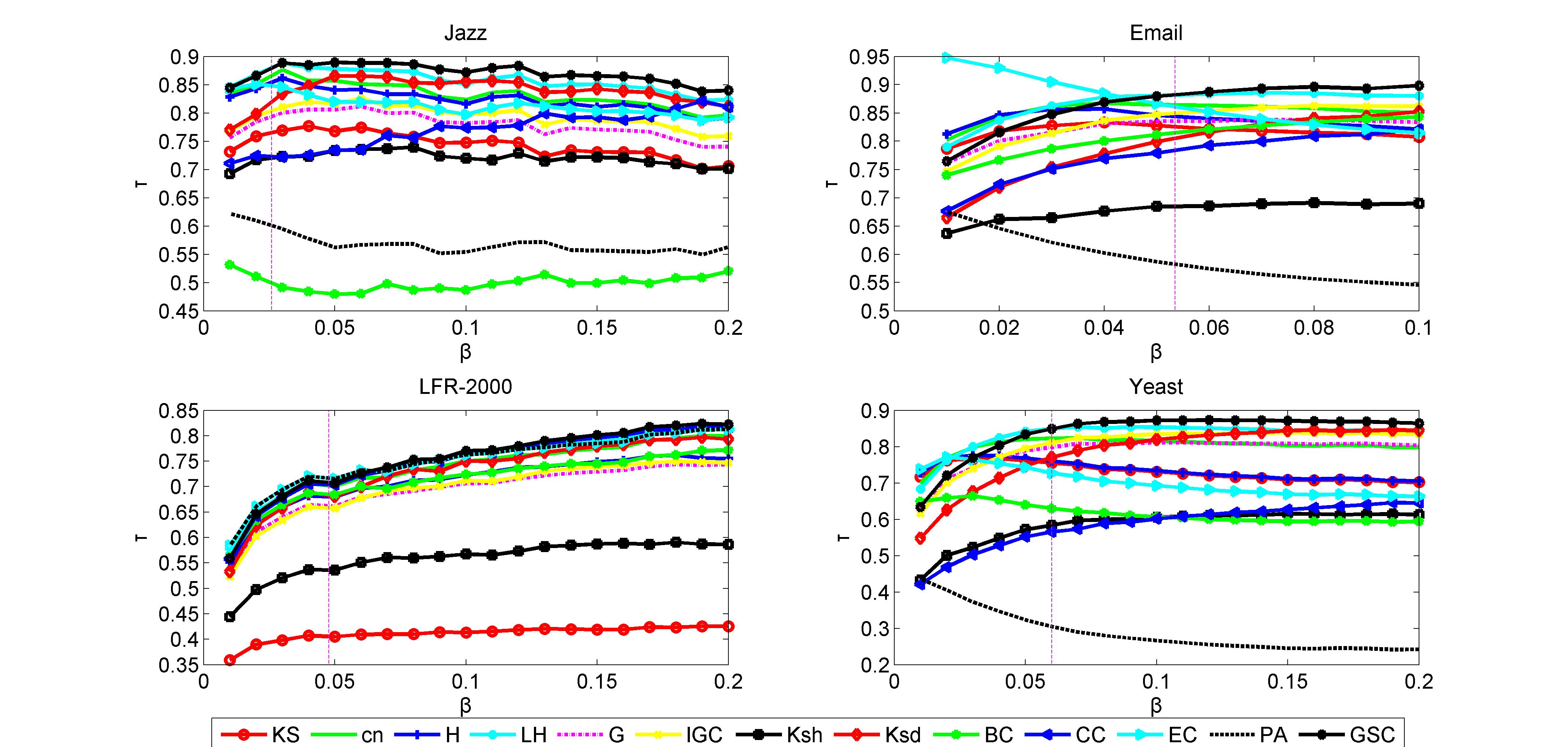}\\
        \caption{The influence of the change of infection rate on the accuracy of different methods in four data sets including Jazz, Email, LFR-2000 and Yeast.}\label{1}
    \end{figure}

\begin{table*}
Table 4: Top-10 nodes ranked by different centrality methods in five real-world networks and the simple graph network.
\centering
\begin{tabular}{ccccccccccc} 
\toprule
\multirow{2}{*}{Rank} & \multicolumn{5}{c}{Karate} & \multicolumn{5}{c}{Dolphins}\\
\cline{2-6}
\cline{7-11}  
   &    KS & cn & H & LH & GSC & KS & cn & H & LH & GSC \\
  \midrule
  1 & $\textcolor{RawSienna}{34}$ & $\textcolor{Blue}{1}$  & $\textcolor{RawSienna}{34}$ & $\textcolor{RawSienna}{34}$ & 1  & 60 & $\textcolor{RawSienna}{15}$ & 52 & $\textcolor{RawSienna}{15}$ & 38\\
  2 & $\textcolor{Red}{33}$ & $\textcolor{RawSienna}{34}$ & $\textcolor{Red}{33}$ & $\textcolor{Blue}{1}$  & 34 & 58 & $\textcolor{Turquoise}{46}$ & $\textcolor{MidnightBlue}{51}$ & $\textcolor{Turquoise}{46}$ & 15\\
  3 & $\textcolor{Dandelion}{31}$ & $\textcolor{Turquoise}{3}$  & $\textcolor{Orange}{14}$ & $\textcolor{Turquoise}{3}$  & 3  & 55 & $\textcolor{Blue}{38}$ & $\textcolor{Turquoise}{46}$ & $\textcolor{Blue}{38}$ & 46\\
  4 & $\textcolor{Orange}{14}$ & $\textcolor{Red}{33}$ & $\textcolor{Turquoise}{3}$  & $\textcolor{Red}{33}$ & 33 & 53 & $\textcolor{Red}{34}$ & $\textcolor{Orange}{41}$ & $\textcolor{Red}{34}$ & 34\\
  5 & $\textcolor{MidnightBlue}{9}$  & $\textcolor{BrickRed}{2}$  & $\textcolor{Blue}{1}$  & $\textcolor{BrickRed}{2}$  & 9  & 52 & 21 & $\textcolor{Blue}{38}$ & 21 & 51\\
  6 & 8  & $\textcolor{Fuchsia}{4}$  & $\textcolor{Dandelion}{31}$ & $\textcolor{Fuchsia}{4}$  & 14 & $\textcolor{MidnightBlue}{51}$ & $\textcolor{Fuchsia}{30}$ & $\textcolor{Red}{34}$ & $\textcolor{Fuchsia}{30}$ & 41\\
  7 & $\textcolor{Fuchsia}{4}$  & $\textcolor{Peach}{32}$ & 24 & $\textcolor{Orange}{14}$ & 32 & 48 & $\textcolor{Orange}{41}$ & $\textcolor{Fuchsia}{30}$ & 52 & 22\\
  8 & $\textcolor{Turquoise}{3}$  & $\textcolor{Orange}{14}$ & $\textcolor{MidnightBlue}{9}$  & $\textcolor{MidnightBlue}{9}$  & 2  & $\textcolor{Turquoise}{46}$ & 52 & 25 & $\textcolor{MidnightBlue}{51}$ & 19\\
  9 & $\textcolor{BrickRed}{2}$  & $\textcolor{MidnightBlue}{9}$  & 8  & $\textcolor{Peach}{32}$ & 4  & 44 & 58 & $\textcolor{Dandelion}{22}$ & $\textcolor{Orange}{41}$ & 30\\
  10& $\textcolor{Blue}{1}$  & 24 & $\textcolor{Fuchsia}{4}$  & 24 & 31 & 43 & 2  & 21 & $\textcolor{BrickRed}{19}$ & 17\\
  \bottomrule
\multirow{2}{*}{Rank} & \multicolumn{5}{c}{Polbooks} & \multicolumn{5}{c}{Football}\\
\cline{2-6}
\cline{7-11}  
   &    KS & cn & H & LH & GSC & KS & cn & H & LH & GSC \\
  \midrule
  1 & 101 & $\textcolor{Blue}{9}$  & $\textcolor{Red}{74}$ & $\textcolor{Blue}{9}$  & 9  & 115 & $\textcolor{Blue}{105}$ & 84 & $\textcolor{RawSienna}{68}$ & 68\\
  2 & 100 & $\textcolor{RawSienna}{13}$ & $\textcolor{Turquoise}{85}$ & $\textcolor{Turquoise}{85}$ & 13 & 114 & $\textcolor{MidnightBlue}{89}$  & 74 & $\textcolor{Red}{54}$ &  8\\
  3 & 92  & $\textcolor{Turquoise}{85}$ & $\textcolor{MidnightBlue}{74}$ & $\textcolor{RawSienna}{13}$ & 85 & 113 & $\textcolor{RawSienna}{68}$  & $\textcolor{RawSienna}{68}$ & $\textcolor{MidnightBlue}{89}$ &  3\\
  4 & 87  & $\textcolor{Orange}{4}$  & 83 & $\textcolor{Red}{74}$ & 74 & 112 & $\textcolor{Red}{54}$  & $\textcolor{Red}{54}$ & $\textcolor{Orange}{16}$ & 54\\
  5 & $\textcolor{Turquoise}{85}$  & $\textcolor{BrickRed}{73}$ & 77 & $\textcolor{MidnightBlue}{31}$ & 31 & 111 & $\textcolor{Orange}{16}$  & 50 & $\textcolor{Turquoise}{3}$  & 89\\
  6 & 84  & $\textcolor{Red}{74}$ & 76 & $\textcolor{BrickRed}{73}$ & 4  & 110 & $\textcolor{BrickRed}{8}$   & 48 & $\textcolor{BrickRed}{8}$  & 16\\
  7 & 83  & $\textcolor{MidnightBlue}{31}$ & $\textcolor{Fuchsia}{75}$ & $\textcolor{Orange}{4}$  & 67 & 109 & $\textcolor{Peach}{7}$   & 47 & $\textcolor{Peach}{7}$  & 105\\
  8 & 80  & $\textcolor{Peach}{67}$ & $\textcolor{BrickRed}{73}$ & $\textcolor{Peach}{67}$ & 73 & 108 & 6   & 33 & $\textcolor{Blue}{105}$&  7\\
  9 & 77  & 48 & $\textcolor{Peach}{67}$ & 76 & 12 & 107 & $\textcolor{Dandelion}{4}$   & $\textcolor{Orange}{16}$ & 2  &  1\\
  10& 76  & 41 & 48 & $\textcolor{Fuchsia}{75}$ & 75 & 106 & $\textcolor{Turquoise}{3}$   & $\textcolor{BrickRed}{8}$  & $\textcolor{Fuchsia}{1}$  &  4\\
  \bottomrule
\multirow{2}{*}{Rank} & \multicolumn{5}{c}{Jazz} & \multicolumn{5}{c}{Simple graph(Fig.1)}\\
\cline{2-6}
\cline{7-11}  
   &    KS & cn & H & LH & GSC & KS & cn & H & LH & GSC \\
  \midrule
  1 & 172 & $\textcolor{Blue}{100}$ & $\textcolor{Blue}{100}$ & $\textcolor{Blue}{100}$ & 100 & 13 & 7  & 13 & 7  &  7\\
  2 & 168 & $\textcolor{RawSienna}{8}$   & $\textcolor{RawSienna}{8}$   & $\textcolor{RawSienna}{8}$   & 8   & 12 & 4  & 12 & 13 &  4\\
  3 & 158 & $\textcolor{Turquoise}{4}$   & $\textcolor{Turquoise}{4}$   & $\textcolor{Turquoise}{4}$   & 4   & 11 & 13 & 11 & 11 & 12\\
  4 & $\textcolor{Red}{131}$ & $\textcolor{Red}{131}$ & $\textcolor{Red}{131}$ & $\textcolor{Red}{131}$ & 131 & 10 & 12 & 10 & 12 & 11\\
  5 & 130 & $\textcolor{MidnightBlue}{80}$  & $\textcolor{Peach}{129}$ & $\textcolor{MidnightBlue}{80}$  & 80  & 8  & 11 & 7  & 10 & 10\\
  6 & $\textcolor{Peach}{129}$ & $\textcolor{Peach}{129}$ & $\textcolor{MidnightBlue}{80}$  & $\textcolor{Peach}{129}$ & 194 & 7  & 10 & 8  & 4  & 13\\
  7 & 106 & $\textcolor{BrickRed}{5}$   & $\textcolor{Fuchsia}{53}$  & $\textcolor{BrickRed}{5}$   & 129 & 6  & 5  & 6  & 8  &  8\\
  8 & 105 & $\textcolor{Tan}{32}$  & $\textcolor{BrickRed}{5}$   & $\textcolor{Orange}{194}$ & 5   & 5  & 8  & 5  & 5  &  6\\
  9 & 104 & $\textcolor{Orange}{194}$ & $\textcolor{Orange}{194}$ & $\textcolor{Fuchsia}{53}$  & 53  & 4  & 6  & 4  & 6  &  1\\
  10& 103 & 84  & $\textcolor{Dandelion}{69}$  & $\textcolor{Dandelion}{69}$  & 69  & 3  & 3  & 3  & 3  &  5\\
  11& 102 & $\textcolor{Dandelion}{69}$  & 130 & $\textcolor{Tan}{32}$  & 162 & 2  & 2  & 2  & 2  &  3\\
  12& $\textcolor{Blue}{100}$ & 85  & 85  & $\textcolor{Plum}{162}$ & 32  & 1  & 1  & 1  & 1  &  2\\
  13& 98  & 53  & 84  & 77  & 59  & 9  & 9  & 9  & 9  &  9\\
  \bottomrule
  \end{tabular}
  \label{tbl:table1}
\end{table*}

\subsection{Similarity}
In the last experiment, disparate measures will generate diverse ranking lists because of considering the different aspects of network topology structure, so we use the number of same high-order vertices in each list to determine the similarity between the methods \cite{47}. The numbers of same nodes increases the credibility of the measure, while the unique nodes in the $GSC$ list will bring significant changes to the spreading process. Experimental results are shown in Table 4. In karate network, $KS$, $cn$, $H$ and $LH$ algorithms have high matching degree with $GSC$ measure, and the number of the same nodes is 9, 9, 8, 9, respectively. In the small-scale networks, the number in the Dolphins network is 2, 6, 7, 8, the number in the Polbooks network is 1, 8, 6, 9, the number in the Football network is 0, 9, 4, 9, and the number in the Jazz network is 3, 10, 10, 12. $KS$ algorithm gradually weakens with the increase of network size, while the other three algorithms are still similar to $GSC$. In the simple graph Fig. 1, compared with the other four algorithms, the proposed measure further details the importance of nodes in the network, and better shows the performance of nodes in the network.


\section{Conclusion}
How to identify and select users to efficiently spread information has become one of the most concerned research topics. In order to achieve this goal, finding the influential nodes is the widely used method. In this paper, a new method is proposed to evaluate the importance of nodes in complex networks: classifying nodes based on distance matrix and combining the correlation between nodes, then applying the global clustering coefficient of networks to the research of node importance. Through extensive experiments on both artificial networks and real-world networks, compared our algorithm with the current popular algorithms, we demonstrate that the the proposed method has a better performance in accuracy, similarity, discrimination capability and other aspects, and which is valuable and significant for the further research.

\medskip


\end{document}